\documentclass[aip,jap,reprint,graphicx]{revtex4-1}
\usepackage{graphicx}% Include figure files
\usepackage{dcolumn}% Align table columns on decimal point
\usepackage{bm}% bold math
\begin{document}
\title{Spin-phonon coupling probed by infrared transmission spectroscopy in the double perovskite Ba$_2$YMoO$_6$}
\author{Zhe Qu}\email[Author to whom correspondence should be addressed. Electronic mail: ]{zhequ@hmfl.ac.cn}\affiliation{High Magnetic Field Laboratory, Chinese Academy of Sciences, Hefei, Anhui, 230031, China}
\author{Youming Zou}\affiliation{High Magnetic Field Laboratory, Chinese Academy of Sciences, Hefei, Anhui, 230031, China}
\author{Shile Zhang}\affiliation{High Magnetic Field Laboratory, Chinese Academy of Sciences, Hefei, Anhui, 230031, China}
\author{Langsheng Ling}\affiliation{High Magnetic Field Laboratory, Chinese Academy of Sciences, Hefei, Anhui, 230031, China}
\author{Lei Zhang}\affiliation{High Magnetic Field Laboratory, Chinese Academy of Sciences, Hefei, Anhui, 230031, China}
\author{Yuheng Zhang}\affiliation{High Magnetic Field Laboratory, Chinese Academy of Sciences, Hefei, Anhui, 230031, China}\affiliation{High Magnetic Field Laboratory, University of Science and Technology of China, Hefei 230026, China}
\begin{abstract}
In this work, we report a study on the infrared transmission spectroscopy of the double perovskite Ba$_2$YMoO$_6$. At 300 K, three bands are observed at $\sim$ 255.1 cm$^{-1}$, $\sim$ 343.4 cm$^{-1}$, and $\sim$ 561.5 cm$^{-1}$, which are related to the motion between the cation Ba$^{2+}$ and the anion YMO$_6^{-2}$, the Y-O stretching motion and the stretching vibration of the MoO$_6$ octahedron, respectively. These modes continue to harden upon cooling owing to the shrink of the lattice constant. When the temperature decreases to $T \leq$ 130 K around which the spin singlet dimer begins to form, an additional phonon mode appears at $\sim$ 611 cm$^{-1}$, suggesting the occurrence of local distortion of MoO$_6$ octahedra. With further decrease of the temperature, its intensity enhances and its peak position keeps unchanged. These results indicate that the formation of the spin singlet dimers is accompanied with the occurrence of the local structure distortion of MoO$_6$ octahedra, providing evidence for the strong spin-phonon coupling in the double perovskite Ba$_2$YMoO$_6$.
\end{abstract}
 \maketitle

%geometry frustration. double perovskite
%which is suggested to be a rare example of the valence bond glass state.
%Meanwhile, it is found that while Sr substitution for Ba MoO$_6$ octahedra induced by . These results seem to suggest that the lattice distortion might not be coupled with the formation of the valence bond glass state this system.

Geometry frustrated magnetic materials have attracted widespread attention in recent decades because of the discovery of exotic ground states such as spin glass, spin ice, and spin liquid. \cite{GFM1,GFM2,GFM3,GFM4} Lattices based on triangular or tetrahedral architectures usually show geometry frustration. This includes the B-site ordered double perovskites with general formula A$_2$BB$^\prime$O$_6$. In these materials, the magnetic ions occupy the B$^\prime$ site while alkaline-earths and lanthanides reside on the B site, forming two interpenetrating lattices with face-centered-cubic (fcc) symmetry. Since the magnetic moments form an edge-sharing tetrahedra network, the magnetism should be geometric frustrated. Considerate efforts have been devoted to experimentally investigate the magnetism of these double perovskites and a wealth of ground states have been revealed. \cite{DBRe1,DBRe2,DBRe3,DBRu1,DBRu2,DBOs1,DBMo1,DBMo2}

Here we focus on the double perovskite Ba$_2$YMoO$_6$, which is an example of the extreme $s =$ 1/2 case and has an ideal cubic double perovskite structure at room temperature.
While the Curie-Weiss temperature determined from bulk susceptibility measurements suggests strong antiferromagnetic (AFM) exchange interaction, no sign of static long-range magnetic order is observed above 2 K. \cite{Ba2YMoO6PRL,Ba2YMoO6PRB} It is found that with decreasing temperature the spins of adjacent $s =$ 1/2 Mo$^{5+}$ gradually freezes into a disordered pattern of spin singlets without signatures of a phase transition, resulting in an exotic ground state with coexisting paramagnetism and collective spin singlets. \cite{Ba2YMoO6PRL,Ba2YMoO6PRB,Ba2YMoO6neutron} Above $\sim$ 125 K, a paramagnetic state is recovered in the system, without obvious signatures of a phase transition. \cite{Ba2YMoO6PRL,Ba2YMoO6neutron} Previous results show that
although Jahn-Teller distortion could be expected to occur owing to $d^1$ electronic configuration of Mo$^{5+}$ ions, Ba$_2$YMoO$_6$ maintains cubic symmetry down to 2 K  and no static structure distortion was observed by neutron diffraction down to 2 K. \cite{Ba2YMoO6PRB,BaSr2YMoO6}

%However, this does excludes the possibility of the occurrence of the local lattice distortions in this compound. Infrared spectroscopy is very sensitive to structural variations, making it suitable for studying the local lattice distortions in complex oxides. Therefore, to better understand the exotic quantum state in Ba$_2$YMoO$_6$, it is instructive to elucidate the coupling between spin and lattice in Ba$_2$YMoO$_6$.

It is known that the coupling between spin and lattice degrees of freedom usually plays an important role in geometry frustrated materials. Infrared spectroscopy is very sensitive tool to probe the spin-phonon coupling. Strong spin-phonon interactions are expected to manifest themselves as either anomalous frequency shifts or occurrence of new phonon peaks with the magnetic transition in infrared spectra. \cite{SPC1,SPC2,SPC3,SPC4,SPC5}

In this work, we measured the infrared transmission spectra of Ba$_2$YMoO$_6$ to explore the spin-phonon coupling in this compound. At 300 K, we observed three bands corresponding to the motion between the cation Ba$^{2+}$ and the anion YMO$_6^{-2}$, the Y-O stretching motion and the stretching vibration of the MoO$_6$ octahedron, respectively. These modes continue to harden upon cooling owing to the shrink of the lattice constant. When the temperature decreases to $T \leq$ 130 K around which the spin singlet dimer begins to form, an additional phonon mode appears at $\sim$ 611 cm$^{-1}$, suggesting the occurrence of local lattice distortion of MoO$_6$ octahedra. With further decrease of the temperature, the intensity of this phonon mode enhances and its peak position keeps almost unchanged. These results indicate that the formation of the spin singlet dimers is accompanied with the occurrence of the local structure distortion of MoO$_6$ octahedra, providing evidence for the strong spin-phonon coupling in the double perovskite Ba$_2$YMoO$_6$.

%In double perovskite, it is known that lattice distortion usually play an important role. Therefore, to better understand the exotic quantum state in Ba$_2$YMoO$_6$, it is instructive to elucidate the lattice distortion in this compound.

\begin{figure}[t]
\includegraphics[angle=-90,scale=1]{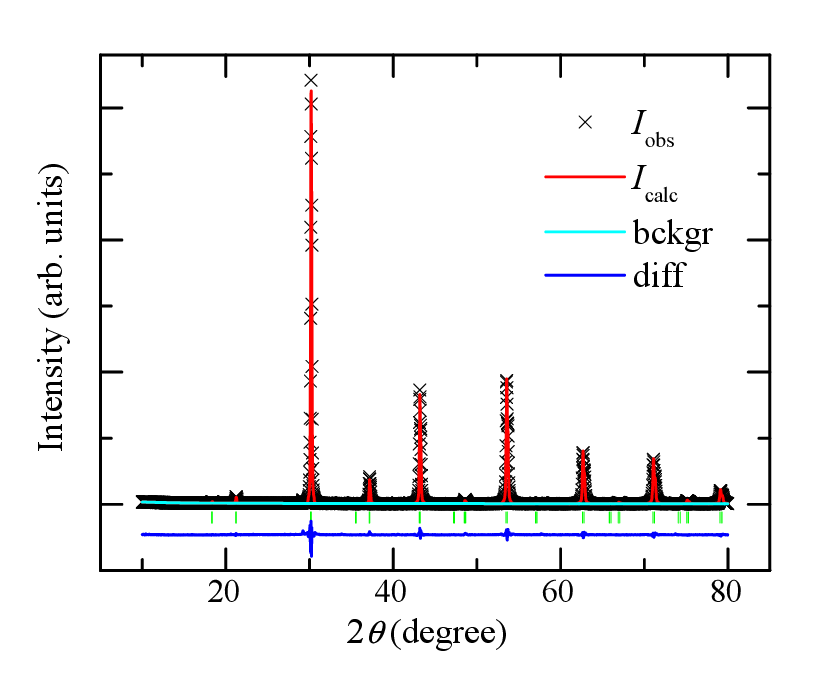}
\caption{(Color online) Powder XRD patterns of Ba$_2$YMoO$_6$. The solid curve is the best fit from the Rietveld refinement using GSAS, with $R_p =$ 5.97\% and $R_{wp} =$ 8.73\%. The vertical marks indicate the position of Bragg peaks and the bottom curves show the difference between the observed and calculated intensities.}\label{fig:XRD}
\end{figure}

\begin{figure}[b]
\includegraphics[angle=-90,scale=1]{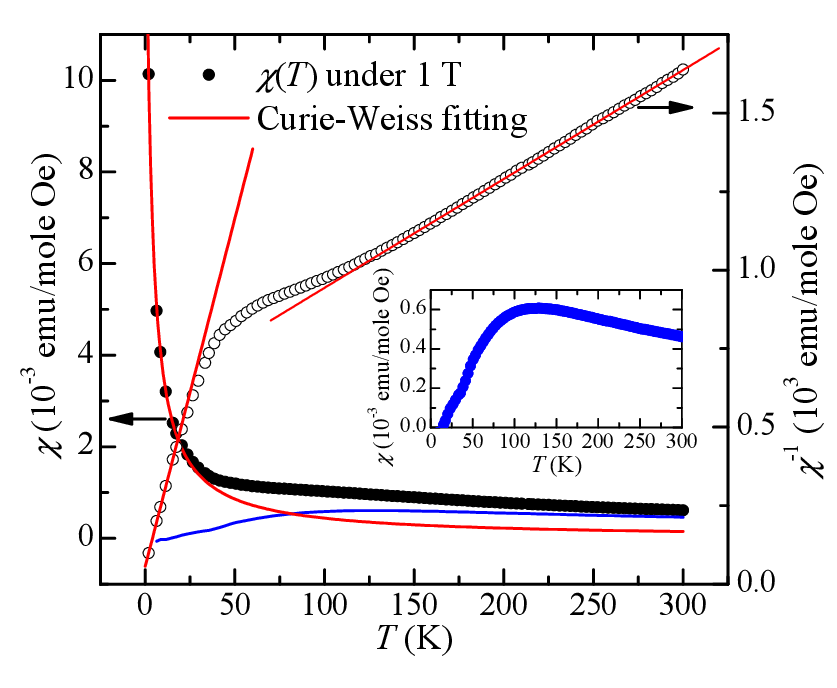}
\caption{(Color online) The susceptibility $\chi$ and its reciprocal $\chi^{-1}$ measured under 1 T as function of the temperature. Red lines represent the Curie-Weiss fitting for the susceptibility in the two linear regions in $\chi^{-1}$. Shown in the inset is the susceptibility after the substraction of the low temperature Curie tail.}\label{fig:Mag}
\end{figure}

\begin{figure}[b]
\includegraphics[angle=0,scale=1]{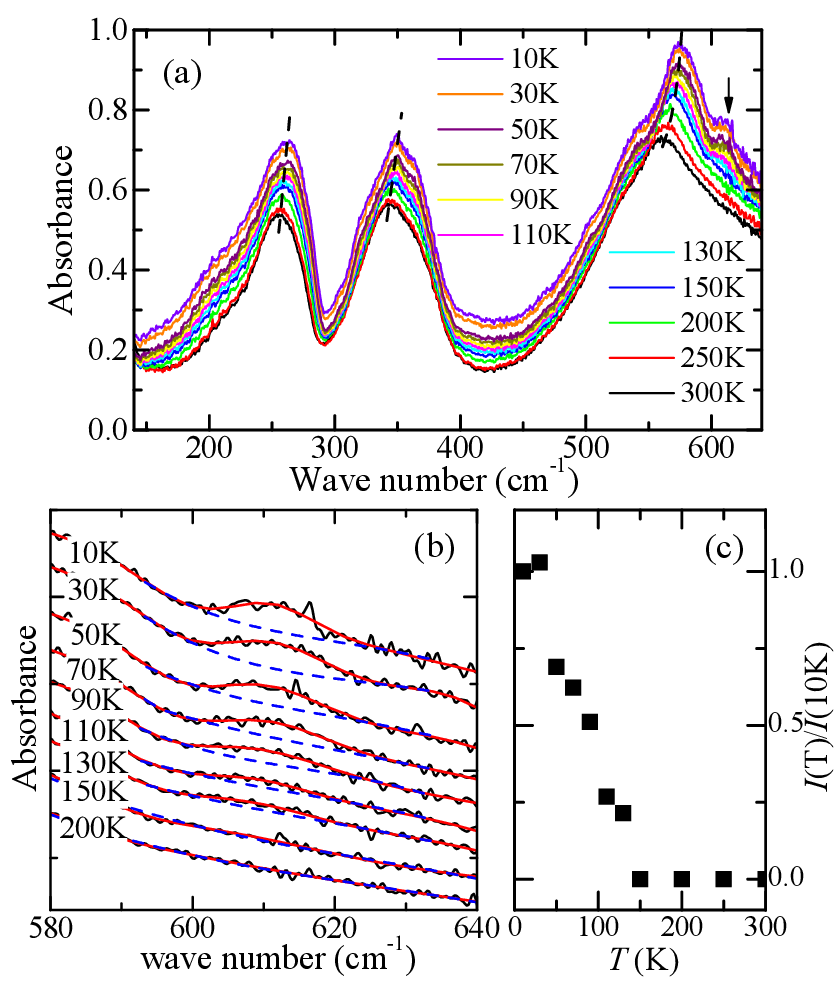}
\caption{(Color online) (a) The infrared transmission spectra measured at various temperatures for Ba$_2$YMoO$_6$ plotted as the absorbance versus the wave number. The dashed lines are guide for the eyes. The arrow marks the additional phonon mode which appears below $\sim$ 130 K. Shown in (b) is the enlarged view of the new phonon mode. The data have been shifted for clarification. The temperature dependence of the normalized intensity of this phonon mode is shown in (c).}\label{fig:IR}
\end{figure}

Polycrystalline Ba$_2$YMoO$_6$ was prepared using the conventional solid-state reaction method. \cite{Ba2YMoO6PRL,Ba2YMoO6PRB} Stoichiometric proportions of high purity Y$_2$O$_3$, MoO$_3$ and BaCO$_3$ powders were mixed and heated at 900 $^{o}$C for 12 hours. They are then pelletized, and then sintered at 1200 - 1250 $^{o}$C under flowing 5 \% H2/Ar gas. The structure and the phase purity of the samples were checked by powder X-ray diffraction (XRD) at room temperature. Magnetization measurements were performed with a commercial superconducting quantum interference device (SQUID) magnetometer (Quantum Design MPMS 7T-XL). The infrared transmission spectra were collected using a Bruker v80v FTIR spectrometer equipped with a Janis continuous flow cryostat (model ST-100-FTIR).

Figure \ref{fig:XRD} displays the powder XRD pattern of Ba$_2$YMoO$_6$ at room temperature. Rietveld refinement \cite{GSAS,EXPGUI} of the XRD pattern confirms that the sample is single phase with a cubic structure ($Fm3m$ space group). The lattice parameter is determined to be $a =$ 8.3853 ${\AA}$, which agrees well with previous reports within the experimental error. \cite{Ba2YMoO6PRL,Ba2YMoO6PRB}

Figure \ref{fig:Mag} shows the temperature dependence of the susceptibility $\chi$($T$) and its reciprocal. It can be seen that there are two regions that follow the Curie-Weiss law. The Curie-Weiss fitting for the susceptibility above 150 K yields a Curie-Weiss temperature  of -172 K, indicating the strong AFM exchange. The effective moment is determined to be 1.52 $\mu_B$, which is smaller than the spin only value for $S =$ 1/2 due to the strong quantum fluctuations \cite{DBOs1,Ba2YMoO6PRL}. Below $\sim$ 25 K, there is another linear regime in $\chi^{-1}$. The Curie-Weiss fitting in this region shows a Curie-Weiss temperature of -2.2 K suggesting the weak AFM interaction. The effective moment is found to be 0.57 $\mu_B$, which corresponds to $\sim$ 10\% fraction of all the $S =$ 1/2 moments. The inset to Fig. \ref{fig:Mag} exhibits the temperature dependence of the susceptibility after the substraction of the low temperature Curie tail. It displays a gap-like feature with a peak at $\sim$ 130 K, which is related to the spin-gap opening at this temperature. \cite{Ba2YMoO6PRB,Ba2YMoO6neutron} All these results are consistent with previous reports, \cite{Ba2YMoO6PRL,Ba2YMoO6PRB,BaSr2YMoO6} confirming that our sample is of high quality.

We further measured the infrared transmission spectra of Ba$_2$YMoO$_6$ at various temperature to probe the lattice distortion in this compound. The results are shown in Fig. \ref{fig:IR} (a). There are three well defined absorption bands associated with the lattice vibrational modes (phonons) having the typical profile expected for an insulating material. At 300 K, these strong phonon modes are observed near $\sim$ 255.1 cm$^{-1}$, $\sim$ 343.4 cm$^{-1}$, and $\sim$ 561.5 cm$^{-1}$, respectively. The phonon band centered at about 255.1 cm$^{-1}$ should be related to the motion between the cation Ba$^{2+}$ and the anion YMO$_6^{-2}$ similar to other double perovskite compounds. \cite{IRDB} The phonon band around $\sim$ 343.4 cm$^{-1}$ can probably related to the Y-O stretching motion, as that observed in Ba$_2$YNbO$_6$. \cite{IRmode} The strong high-energy phonon band located near 561.5 cm$^{-1}$ should be assigned to the stretching vibration mode of the MoO$_6$ octahedron due to the higher charge of this cation. \cite{IRmode}

With decreasing temperature, one can see that all these three bands are displaced toward higher energy direction, implying an enhancement of the bond strength. This is in agreement with the slight shrinkage of the lattice constant upon cooling in Ba$_2$YMoO$_6$. \cite{Ba2YMoO6PRB,BaSr2YMoO6}
More importantly, as shown in Fig. \ref{fig:IR} (b), an additional phonon mode occurs at $\sim$ 611 cm$^{-1}$ when the temperature decreases to $T \leq$ 130 K. This fact suggests that a new stretching vibration mode of the MoO$_6$ octahedron appears below $\sim$ 130 K, hinting that a local distortion of MoO$_6$ octahedron should occur. Since the occurrence temperature of this mode corresponds to the peak temperature in $\chi$($T$) curve around which the spin-singlet pairs begin to form and the spin gap opens,\cite{Ba2YMoO6PRB,Ba2YMoO6neutron} the occurrence of such phonon peak suggests that there is strong spin-phonon coupling in Ba$_2$YMoO$_6$. 
This argument is further supported by the temperature dependence of this additional phonon peak.
It can be seen that its peak position keeps almost unchanged with further decrease of the temperature, hinting that the strength of the local lattice distortion does not change. 
On the other hand, the intensity of this phonon mode enhances upon cooling and approaches saturation below $\sim$ 30 K (see Fig. \ref{fig:IR} (c)), which means that more and more MoO$_6$ octahedra become distorted. This results can be understood by considering the magnetic properties of Ba$_2$YMoO$_6$; since increasing numbers of spin singlets are formed upon cooling, the local lattice distortion is expected to occur in increasing number of MoO$_6$ octahedra.
All these results indicate that the formation of the spin singlet dimers is accompanied with the occurrence of the local structure distortion of MoO$_6$ octahedra, suggesting that there is strong spin-phonon coupling in the double perovskite Ba$_2$YMoO$_6$.

%The local lattice distortion observed here should be closely related to the formation of spin singlet dimers in Ba$_2$YMoO$_6$, since spins of adjacent $s =$ 1/2 Mo$^{5+}$ begin to pair into spin singlet dimers in Ba$_2$YMoO$_6$ at the same temperature where the local lattice distortion occurs. \cite{Ba2YMoO6PRL,Ba2YMoO6PRB,Ba2YMoO6neutron}

%conclusion

In summary, we have conducted an infrared transmission spectroscopy study in the double perovskite Ba$_2$YMoO$_6$ to investigate its spin-phonon coupling. Three phonon bands have been identified at 300 K. We find that an additional phonon mode appears at $\sim$ 611 cm$^{-1}$ when the temperature decreases to $T \leq$ 130 K around which the spin singlet dimer begins to form, suggesting the occurrence of local distortion of MoO$_6$ octahedra. With further decrease of the temperature, its intensity enhances and its peak position keeps unchanged. Our results indicate that the formation of the spin singlet dimers is accompanied with the occurrence of the local structure distortion of MoO$_6$ octahedra, providing evidence for the strong spin-phonon coupling in the double perovskite Ba$_2$YMoO$_6$.

%Ba$_2$YMoO$_6$

\begin{acknowledgments}
We thank Drs. Zhiquan Jiang, Ranran Zhang, Jun Fang, and Wei Tong for helpful discussions. This work is supported by National Natural Science Foundation of China under contracts Nos. 11004198 and 11174291. Z. Q. gratefully acknowledges supports from the Youth Innovation Promotion Association, Chinese Academy of Sciences.
\end{acknowledgments}

\end{document}